\documentclass[aip,reprint,floatfix]{revtex4-1}

\usepackage{bm}
\usepackage{url}
\usepackage{times}
\usepackage{verbatim}
\usepackage{float}
\usepackage{theorem}
\usepackage{makeidx}
\usepackage{amsmath}
\usepackage{siunitx}
\usepackage{soul}
\usepackage[normalem]{ulem}
\usepackage{fancybox}
\usepackage[english]{babel}
\usepackage{overpic} 
\usepackage{xy}
\usepackage{amssymb}
\usepackage{graphicx}
\usepackage{color}
\usepackage{xcolor}
\usepackage{epsfig}
\usepackage{subcaption}
\usepackage{multirow}
\usepackage[utf8]{inputenc}
\usepackage{mathtools}
\usepackage{array}
\renewcommand{\arraystretch}{1.2}

\newcommand{\br}{\boldsymbol{r}}

\newcommand{\bX}{\boldsymbol{X}}
\newcommand{\Ra}{\rm{Ra}}
\newcommand{\Nu}{\rm{Nu}}
\newcommand{\Tp}{T^i}
\newcommand{\uu }{\mathbf{u}}
\newcommand{\UU }{\bm{\mathcal{U}}}
\newcommand{\TT }{\mathcal{T}}
\newcommand{\nab }{\mathbf{\nabla}}

\usepackage{todonotes}

\usepackage{hyperref}   
\newcommand{\utov}{Department of Physics, University of Rome ``Tor Vergata" and INFN, Via della Ricerca Scientifica 1, 00133, Rome RM, Italy}
\newcommand{\buw}{Angewandte Mathematik und Numerische Analysis, Bergische Universit\"{a}t Wuppertal, Gaußstrasse 20 D-42119 Wuppertal, Germany}
\newcommand{\cyi}{Computation-based Science and Technology Research Center, The Cyprus Institute, 20 Kavafi Str., Nicosia 2121, Cyprus}
\newcommand{\jhu}{Department of Mechanical Engineering, Johns Hopkins University, Baltimore, MD 21218, USA}
\newcommand{\uba}{Departmento de Física, Facultad de Ciencias Exactas y Naturales, Universidad de Buenos Aires, Buenos Aires 1428, Argentina}

\begin{document}

\preprint{AIP/123-QED}

\title{Reconstructing Rayleigh-B\'enard flows out of temperature-only measurements using nudging}
\author{Lokahith Agasthya}
\email{lnagasthya@gmail.com}
\affiliation{\utov}
\affiliation{\buw}
\affiliation{\cyi}

\author{Patricio Clark Di Leoni}%
\affiliation{\jhu}
\affiliation{\uba}

\author{Luca Biferale}
 \homepage{https://people.roma2.infn.it/~biferale/}
\affiliation{\utov}%

\date{\today}

\begin{abstract}
Nudging is a data assimilation technique that has proved to be capable of reconstructing several highly turbulent flows from a set of partial spatiotemporal measurements. In this study we apply the nudging protocol on the temperature field in a Rayleigh-B\'enard Convection system at varying levels of turbulence. We assess the global, as well as scale by scale, success in reconstructing the flow and the transition to full synchronization while varying both the quantity and quality of the information provided by the sparse measurements either on the Eulerian or Lagrangian domain. We asses the statistical reproduction of the dynamic behaviour of the system by studying the spectra of the nudged fields as well as the correct prediction of the heat transfer properties as measured by the Nusselt number. Further, we analyze the results in terms of the complexity of the solutions at various Rayleigh numbers and discuss the more general problem of predicting all state variables of a system given partial or full measurements of only one subset of the fields, in particular temperature. This study sheds new light on the correlation between velocity and temperature in thermally driven flows and on the possibility to control them by acting on the temperature only.
\end{abstract}

\maketitle

\section{Introduction}

{Thermally} driven flows are of fundamental importance in several geophysical as well as industrial processes, including but not limited to atmospheric convection\cite{hartmann2001tropical,suselj2019unified}, convection in the ocean \cite{vaage2018ocean}, mantle convection \cite{kronbichler2012high} and melting of pure metals \cite{brent1988enthalpy}.
The intensity of the thermal instabilities in these flows is characterized by the Rayleigh number, which is the ratio between the buoyant and viscous forces scaled by the ratio of the momentum and thermal diffusivity. At increasing Rayleigh number, natural convection displays a rich set of different dynamical behaviour, including transition to chaos, formation of patterns and finally fully developed turbulence \cite{Grossman-Lohse-Scaling}.
Despite being omnipresent in nature, several facts about convective flows remain unsettled. For example, in Rayleigh-B\'enard convection, it is conjectured that there exists a cross-over length $L_B$, the so-called Bolgiano lengthscale, such that for $L \gg L_b$ the Bolgiano-Obukhov (BO59) \cite{bolgiano1959,obukhov1959} scaling is expected while for $L \ll L_b$ , the scaling is thought to comply with Kolmogorov (K41) prediction \cite{kolmogorov1941,lohse2010small}. The two regimes entail very different velocity-temperature correlations: in the BO59 case temperature is strongly active at all scales and kinetic energy cascades to the small scales via interactions with the thermal component, while in K41, kinetic energy is mainly injected by buoyancy in the bulk and transferred to high wavenumbers via the nonlinear Navier-Stokes terms. Similarly, the interplay between large-scale thermal plumes and small-scale and strongly intermittent fluctuations is also not entirely understood \cite{frisch_turbulence:_1995}. Thus, the exact nature of the expected energy spectrum and the particular scaling regimes is still a subject of active research.


Related to the question of the nature of the velocity-temperature correlations, lies an old conjecture stated by Charney \cite{ch03600a}, which has strong implications to both our fundamental understanding of convective flows and to their applications. The conjecture states that temperature measurements alone are sufficient to predict all the state variables of the atmosphere. 
Early studies \cite{ghil_balanced_1977,ghil_time-continuous_1979} showed that the conjectures is true in simple systems, but also presented numerical evidence to the contrary. In the past few years the conjecture has been studied in different models, such as in a 3D Planetary Geostrophic model \cite{farhat2016charney}, a Rayleigh-B\'enard flow in the infinite Prandtl number limit \cite{farhat2020data}, and a Rayleigh-B\'enard flow in the non-turbulent regime \cite{altaf_downscaling_2017}. The first two studies were successful in reconstructing the whole state out of thermal measurements, but the third study struggled to do so when the prior on the velocity field was not set correctly. So while the conjecture seems to hold in simple systems, it is not clear what can happen in highly-turbulent scenarios with large scale separation. In these cases the correlations between temperature and velocity at the smaller scales may not be strong enough for the temperature to fully enslave the evolution of the velocity field. The conjecture has started to gain even more importance in past years, as weather forecast centers are starting to be able to resolve scales where three-dimensional turbulent convection becomes relevant \cite{Leutwyler16,Yano18}. Outside the realm of convective flows, the question of what fields or components carry enough information to allow for full reconstruction of an entire system is an open-ended problem that sits at the heart of all disciplines related to Data Assimilation (DA) \cite{Kalnay,Bauer15}.

\begin{figure*}
    \centering
    \begin{overpic}[width = \linewidth]{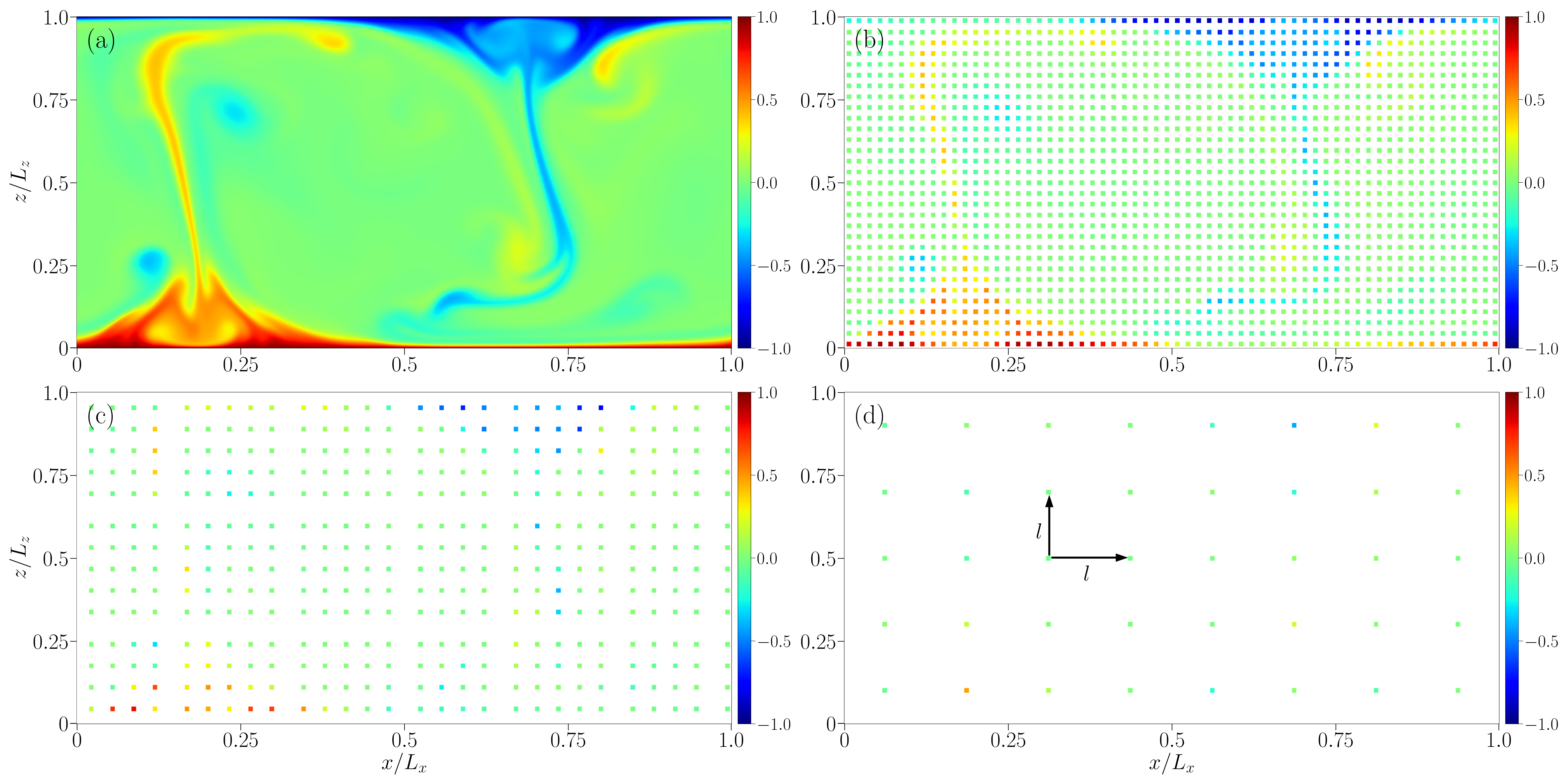}
    \end{overpic}
    \caption{Snapshots of the temperature fields for the reconstruction experiments for flow with $\Ra = 7.2 \times 10^7$. Panel (a) shows the ground-truth while (b), (c), and (d) show the constructed nudging field $T_n(\br,t)$ with $k_l = 1/14$, $k_l= 1/31$ and $1/97$, respectively. The temperatures are normalised to range between $-1$ (blue) and $+1$ (red) by dividing by $T_d$. The individual squares here correspond to the nudging squares $S_i$ with side length $\chi = 6$. The typical length $l$ between the probes is shown in panel (d).}
    \label{fig:Nudging_fields}
\end{figure*}

The main aim of this paper is to address the above issues in turbulent flows head on. We focus on Rayleigh-B\'enard flows in 2D and utilize a Data Assimilation technique called Nudging \cite{Hoke76,Lakshmivarahan13} as a means to reconstruct the whole state of the system given only information on temperature. Nudging consists of adding a relaxation feedback term to the system that penalizes the flow when it strays away from a reference set of data. When the nudged flow synchronizes to the reference flow, missing data from either missing variables or missing scales can be recovered. References \cite{farhat2016charney, farhat2020data, altaf_downscaling_2017} listed above, used nudging to reconstruct the velocity fields using only thermal data, as per the Charney conjecture. In a similar vein, \cite{farhat_continuous_2015,farhat_continuous_2017} studied the problem of nudging a Rayleigh-B\'enard flow using only velocity information from a purely theoretical approach and \cite{farhat_assimilation_2018} did it using only vorticity information. In non-thermal flows, \cite{Lalescu13,clark_di_leoni_synchronization_2020} demonstrated that nudging can accurately reconstruct missing scales in highly turbulent flows, and \cite{clark_di_leoni_inferring_2018, buzzicotti_synchronizing_2020} showed that nudging is sensitive to discrepancies in parameters between the nudged and reference model. Nudging has also been used in finite dimensional dynamical systems and weather models \cite{Hoke76,Auroux08,Du13,Pazo16}, and for boundary condition matching \cite{Vonstorch00,Waldron96,Miguez-macho04}.
Our focus in this paper is on turbulent 2D Rayleigh-B\'enard flows, where we performed numerical experiments to 
investigate the effects of using different amounts of information (i.e., distance between measuring probes), different types of information (whether the data is taken from moving Lagrangian probes or fixed Eulerian ones) and the dependence of the results on the Rayleigh number (i.e., the intensity of the turbulence). The study is a step towards understanding the feasibility of data assimilation techniques on real-world observational data when information on one or more state variables are missing. Our main result concerns the limit up to which the velocity field can be inferred from the temperature field. In particular, we show that there exist a {\it transition } for some characteristic Rayleigh number, $\Ra \sim 10^7$ where the capability to construct the whole velocity configuration deteriorates suddenly.

This article is organised in the following way. In Sec.~\ref{sec:Methods} we give an overview of the nudging technique and the specific protocol employed in this study. Sec.~\ref{sec:Set-up} details the numerical experiments and the parameter space explored. Sec.~\ref{sec:Results} describes the key results of the study. Sec. \ref{sec:Concl} presents our conclusions and future research directions.

\section{Nudging the Rayleigh-B\'enard flow}\label{sec:Methods}

\begin{figure*}
    \centering
    \begin{overpic}[width = \linewidth]{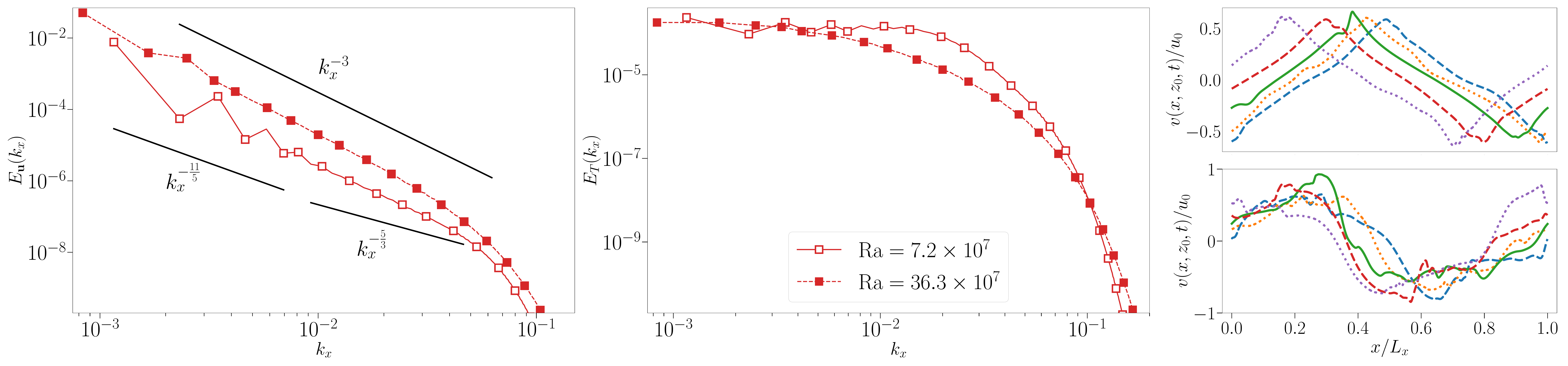}
    \end{overpic}
    \caption{(a) $E_\uu$ and (b) $E_T$ for the two reference flows. Panel (c) shows $v(x,z_0,t)$ at various times during the run for the lower $\Ra$ reference flow with $\Ra = 7.2 \times 10^7$ where each curve represents a particular instant of time. Panel (d) shows the same for the higher $\Ra$ reference flow with $\Ra = 36.3 \times 10^7$. }
    \label{fig:Refs_comp}
\end{figure*}

\begin{figure}
    \centering
    \includegraphics[width = \columnwidth]{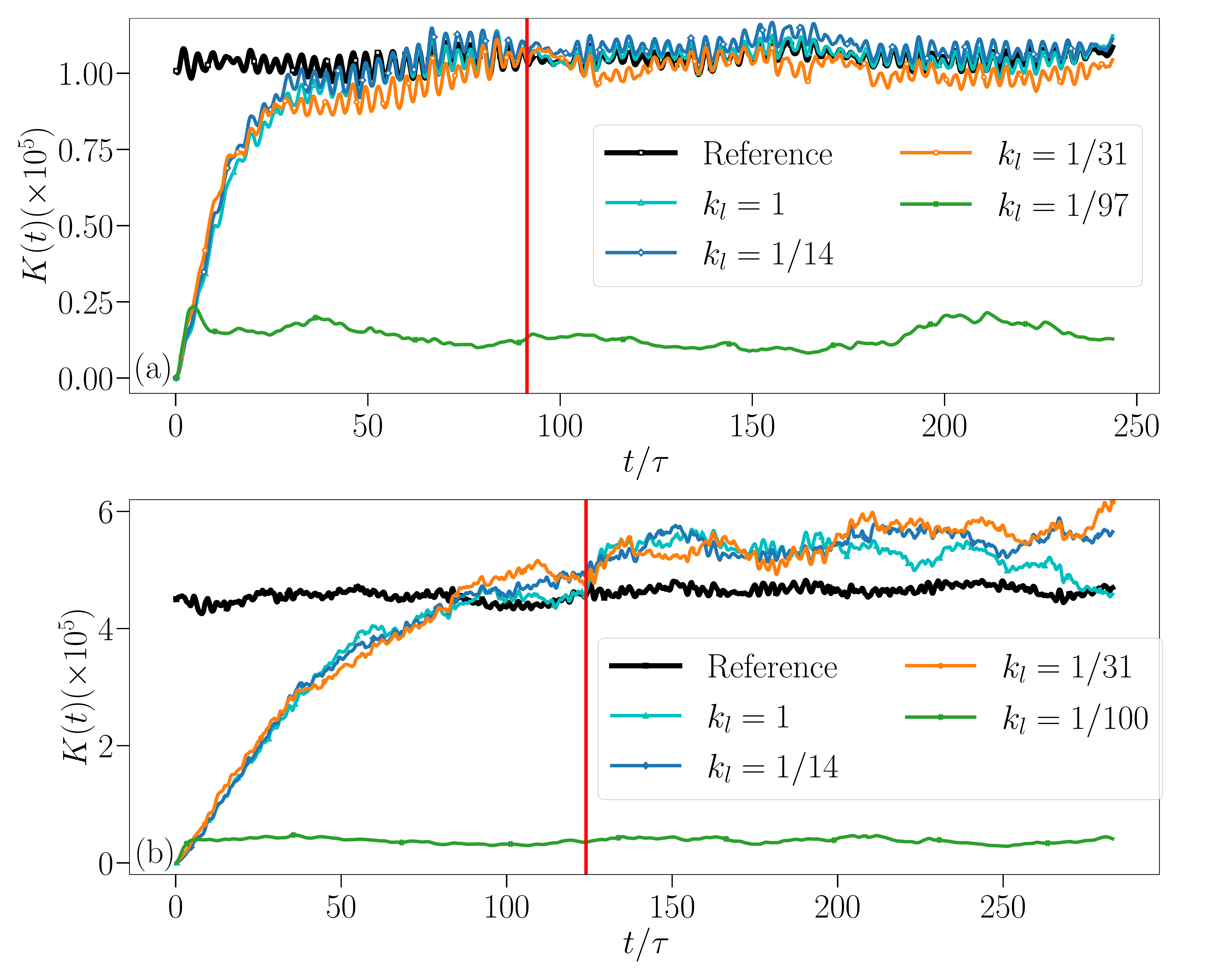}
    \caption{(a) Evolution of the kinetic energy $K(t) (\times 10^5)$ in simulation units for the reference flow with $\Ra = 7.2 \times 10^7$ and the corresponding reconstructed flows for various $k_l$. (b) Evolution of the kinetic energy $K(t)(\times 10^5)$ in simulation units for the Reference flow with $\Ra = 36.3 \times 10^7$ and the corresponding reconstructed flows for various $k_l$. In both panels, the vertical red line shows the time when the reconstructed flows are considered to have attained the statistically stationary regime in which the various measurements and analysis are performed.}
    \label{fig:ke_evol}
\end{figure}

A Rayleigh-B\'enard flow can be described as a compressible fluid bounded at the top and bottom by two plates, each at a specific and constant temperature. Under the Boussinesq approximation, density fluctuations are assumed to be small, such that the flow can be solved in the incompressible regime and density enters the equations only through the buoyancy term. The full 2D equations for the velocity $\uu = (u,v)$, temperature $T$, and pressure $p$ take the form

\begin{gather}\label{eq:R-B-eqns}
    \partial_t \uu + (\uu \cdot \nab) \uu  =  -\nab p  + \nu \nabla^2 \uu - \beta T  \bf g, \\   \label{eq:R-B-eqns1}
    \frac{\partial T}{\partial t} + \uu \cdot \nab T = \kappa \nabla^2 T,
    \\\label{eq:R-B-eqns2}
    \nabla \cdot \uu = 0,
\end{gather}
where the average temperature of the fluid is set to zero and the density to unity. Here, $\beta$ is the thermal expansion coefficient of the fluid at temperature $T_0 = 0$, $\nu$ is the kinematic viscosity of the fluid, $\kappa$ is the thermal conductivity of the fluid and ${\bf g}$ is the acceleration due to gravity. The domain is a rectangle with horizontal and vertical length $L_x$ and $L_z$, respectively, and is periodic in the horizontal $x$ direction. The vertical temperature and velocity boundary conditions are given, respectively, by 

\begin{equation}\label{eq:R-B_BoundaryCs}
\begin{gathered}
   T(z=0) = T_d,\qquad
   T(z=L_z) = -T_d, \\
   \uu(z=0) = \uu(z=L_z) = 0. \\
\end{gathered}
\end{equation}
where $T_d$ is positive. We also define a velocity scale $u_0 = \sqrt{|{\bf g}| L_z \beta \Delta T}$, where $\Delta T = 2 T_d$ is the temperature difference between the top and bottom walls, and a turnover time scale $\tau_0 = 2 L_z/u_0$. The two dimensionless numbers that characterize the flow are the Rayleigh number $\Ra$ and the Prandtl number $\rm Pr$ defined, respectively, as

\begin{equation}
    \begin{gathered}
        {\Ra} = |{\bf g}| \beta \frac{\Delta T L_z^3}{\nu \kappa}, \qquad
        {\rm Pr} = \frac{\nu}{\kappa}.
    \end{gathered}
\end{equation}

An important response parameter measured in the Rayleigh-B\'enard system is the dimensionless Nusselt number, given by 

\begin{equation}\label{eq:Nu-defn}
    {\Nu} (z) = \frac{\langle v T - \kappa \partial_z T \rangle_{x,t} }{\frac{\kappa \Delta T}{L_z}},
\end{equation}
where $\langle \cdot \rangle_{x,t}$ indicates the time and spatial averages  at a given height $z$. The Nusselt number measures the ratio of vertical heat transfer due to convection to the vertical heat transfer due to conduction. In the Rayleigh-B\'enard convection, the Nusselt number is constant and has the same value at all heights so,
\begin{equation}
    \langle {\rm{Nu}} \rangle = {\rm{Nu}}(z), \qquad \text{for } 0 \leq z \leq L_z,
\end{equation}
where $\langle \cdot \rangle$ indicates the average over the entire domain and time.

We also define the Kolmogorov length scale as 
$
    \eta_\kappa = \left( {\nu^3}/{\epsilon} \right)^{\frac{1}{4}}$,
where $\epsilon = ({\nu \kappa^2}/{L_z^4}) \Ra (\langle \Nu \rangle-1)$ is the the average rate of energy dissipation  \cite{siggia1994high}.  

In this study, we assume to have a set of passive probes at positions $\bX^i(t)$, with  $i=1,\dots, N_p$, suspended in a reference Rayleigh-B\'enard flow. The probes make periodic measurements of the fluid temperature $\Tp(t) = T(\bX^i(t),t)$ with constant sampling frequency $f = 1/\tau$ where $\tau$ is the time difference between two successive temperature measurements by the probes. We set $f$ high enough so that a simple linear interpolation in time results in an error smaller than $1$ part in $10^5$. This is done by considering the evolution of the temperature at individual grid points and performing a linear interpolation in time with varying sampling frequency. $f$ was chosen as the smallest value so the sampling was smooth and faster than the fastest time scale of the system. This way additional complexities of errors arising from temporal interpolation are avoided and we can assume to know $\Tp(t)$ at every time-step. In the main section of the paper, the probes are considered to be Eulerian, i.e. their locations $\bX^i(t)$ are kept fixed in the laboratory reference frame. In Appendix B, we discuss and detail the case of Lagrangian probes which do not remain fixed in time but follow trajectories of tracers particles in the flow instead. 

The idea behind nudging is to run a new system where we steer the evolution of the flow onto the path set by the data. To do this, we construct a {\it nudging field} $T_n(\br,t)$ using the measurements $\Tp(t)$ and run another Rayleigh-B\'enard system, denoted by velocity $\UU=(\mathcal{U}, \mathcal{V})$ and temperature $\TT$, where an extra {\it heat source} term  $ -\alpha (\TT - T_n)$ is added to the the evolution equation for $\TT$. This term penalizes the nudged temperature when it deviates from the reference values - if the temperature is higher (lower) than the nudging field, you locally absorb (release) heat. For each probe, we define a nudging square $S_i$ with centre at $\bX^i$ and fixed side length $\chi$ typically chosen of the same order as $\eta_\kappa$, (see also Table \ref{tab:Recons_params}) and we nudge only in regions belonging to the sub-domain $\mathcal{S} \coloneqq \cup_{i = 1}^{N_p} S_i$. We do this by defining $\alpha$ as

\begin{equation}
    \alpha(\br,t) = 
    \begin{cases}
    \alpha_0, & \text{for } \br \in \mathcal{S},\\
    0, & \text{otherwise,}
    \end{cases}
\end{equation}
and the nudging field is defined as 
\begin{equation}
    T_n(\br,t) = \Tp(t), \qquad \text{for } \br \in S_i,
\end{equation}
that is, the measured temperature $\Tp(t) = T(\bX^i(t),t)$ is set as the nudging temperature uniformly in a square of length $\chi$ centred at the probe-location. Thus, the full equations for the nudged fields read 

\begin{gather} \label{eq:R-B-nudged}
  \partial_t \UU + (\UU \cdot \nabla) \UU = -\nab \mathcal{P} + \nu \nabla^2 \UU - \beta \TT \bf{g}, \\
  \frac{\partial \TT}{\partial t} + \UU \cdot \nab \TT = \kappa \nabla^2 \TT - \alpha (\TT - T_n) , \label{eq:T-nudged}
  \\\label{eq:R-B-nudged1}
  \nab \cdot \UU = 0 ,
\end{gather}
where $\alpha(\br,t)$ sets the strength of the coupling between the reconstructed flow and the nudging field and has dimension $1/t$. It is very important to stress that the only energy input into the nudged system is via the nudging field, i.e. we impose adiabatic boundary conditions at the top and bottom walls.

\begin{equation}
    \partial_z \TT |_{z=0}  =
    \partial_z \TT |_{z=L_z} = 0.
\end{equation}

In other words, the basic line for the reconstruction has no prior - if $T_{n}=0$, the flow is zero everywhere. Since the Rayleigh number of the Rayleigh-B\'enard convection is set by the temperature boundaries, it is likely that sampling preferentially from near the walls near the thermal boundary would result in the most accurate reconstruction. However, in the interest of a fair assessment of the nudging protocol, we consider only temperature probes on a uniform array without supposing we know anything about the nature of the thermal boundary. As a result the only information about the $\Ra$ available is that which is encoded in the probe measurements.
    
Here and hereafter we characterize the spatial density of the probes by a characteristic wavenumber  $k_l$ given by 
\begin{equation}
    k_l = \frac{\Delta r}{l},
\end{equation}
where $\Delta r = 1$ is the grid-spacing in the numerical algorithm to evolve the flow and $l$ is the typical distance between the nearest probes. As the number of probes increases, $l$ decreases and $k_l$ increases as $\sqrt{N_p}$. The case of $k_l = 1$ corresponds to the situation where we have complete information, since there are probes located at every grid-point. In Fig.~\ref{fig:Nudging_fields} we show the temperature snapshot of a temperature field as well as the nudging fields $T_n(\br,t)$ constructed from this field for three different values of $k_l$. From the figures, it is clear that for $k_l \gtrsim 0.1$ the density of point measurements should be enough to be able to accurately interpolate the temperature field while for $k_l \lesssim 0.03$  or smaller, the input information is rather limited. The problem then is two-fold --- to understand how much information about the temperature field is needed to reconstruct it to a given degree of accuracy and to investigate the extent up to which the velocity field can be reconstructed from temperature data. 

\section{Numerical Experiments}
\label{sec:Set-up}
\begin{figure*}
    \begin{overpic}[width = \linewidth,keepaspectratio]{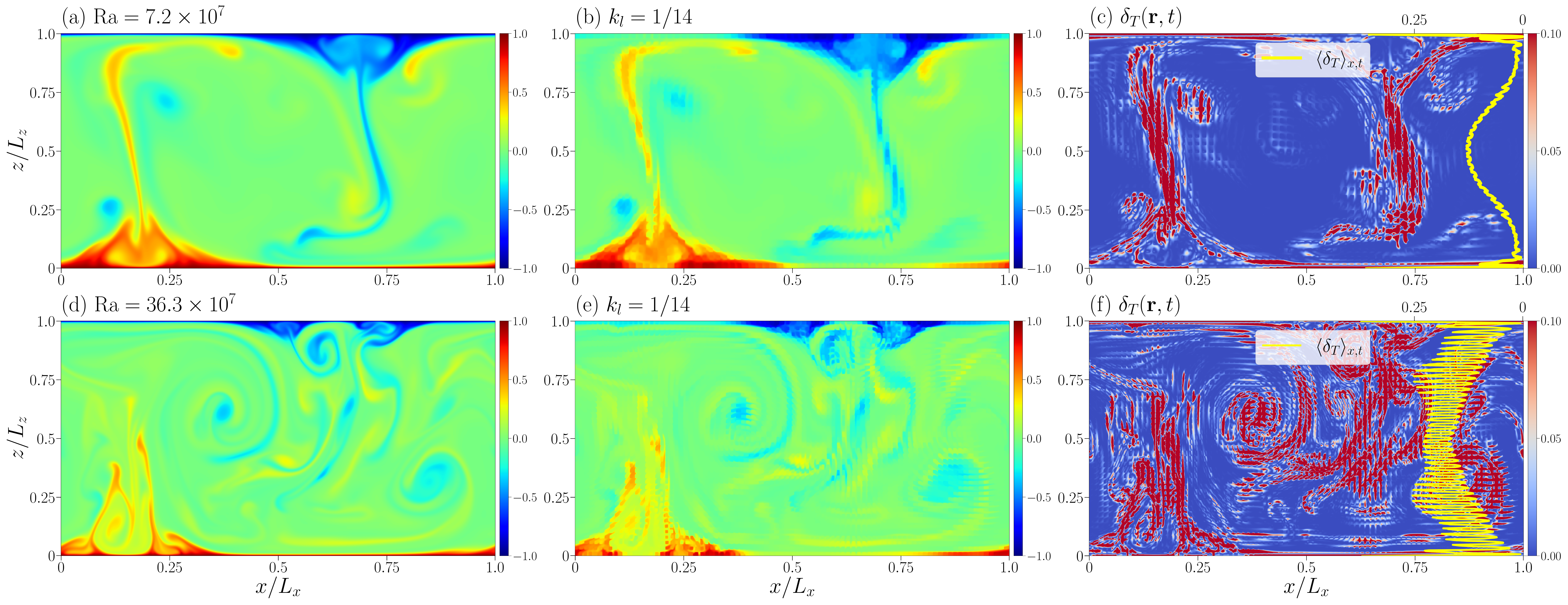}

    \end{overpic}
    \caption{Snapshots of the (a) normalised ground truth field $T/T_d$ for $\Ra = 7.2 \times 10^7$, (b) the normalised reconstructed temperature field, $\mathcal{T}/T_d$ for $k_l=1/14$ corresponding to a distance between probes $l \sim 7 \eta_\kappa$ and (c) the error, $\delta_T(\br,t)$ at a given instant of time. The yellow curve in panel (c) shows the time-averaged vertical profile of $\delta_T$ with the scale shown on the top right. The lower panels show the same quantities for the higher $\Ra$ flow with $\Ra = 36.3 \times 10^7$. }
    \label{fig:Confs_comp_T}
\end{figure*}
\begin{figure*}
    \centering
    \begin{overpic}[width = \linewidth,keepaspectratio]{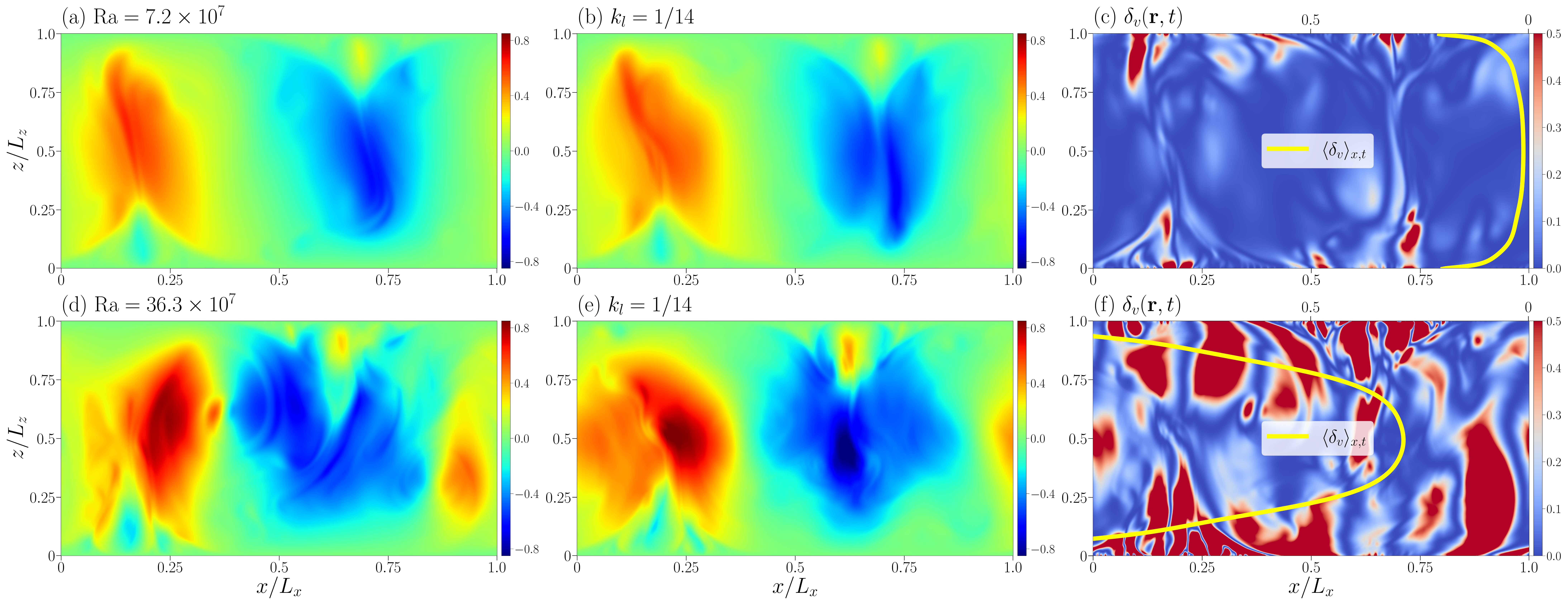}

    \end{overpic}
    \caption{Snapshots of the (a) normalised ground truth field $v/u_0$ for $\Ra = 7.2 \times 10^7$, (b) the normalised reconstructed field, $\mathcal{V}/u_0$ for $k_l=1/14$ and (c) the error, $\delta_v(\br,t)$ at a given instant of time. The yellow curve in panel (c) shows the time-averaged vertical profile of $\delta_v$ with scale shown on the top right. The lower panels show the same quantities for the higher $\Ra$ flow with $\Ra = 36.3 \times 10^7$. }
    \label{fig:Confs_comp_v}
\end{figure*}

\subsection{Reference flows}

Equations ~\eqref{eq:R-B-eqns} - \eqref{eq:R-B-eqns2} are evolved using the Lattice-Boltzmann method (details in Appendix A) until the average kinetic energy of the flow becomes statistically stationary. Once a stationary state is reached, the $N_p$ passive probes are initialised in the domain on a uniform and equally spaced grid and the data from the probes, that is the fluid temperatures $\Tp(t)$ at $\bX^i(t)$, is obtained for $\sim 250$ turn-over times $\tau_0$. This measured data is used to construct the nudging field $T_n$ as detailed in section \ref{sec:Methods}. The flows studied have $\rm Pr = 1$ and $\Ra$ between $10^7$ - $10^9$, which is the well-known transition to turbulence regime in Rayleigh-B\'enard convection. Two flows in particular, one with a moderate value of $\Ra= 7.2 \times 10^7$ and a second one with a higher value of $\Ra = 36.3 \times 10^7$  (see Table. \ref{tab:Ref_params}) are studied and presented in more detail.

\begin{table}
    \centering
    \begin{tabular}{|c|c|c|}
        \hline
        \multicolumn{3}{|c|}{\textbf{Reference Flows}}                        \\
        \hline
        $\boldsymbol{\Ra}$   & $\boldsymbol{7.206 \times 10^7}$ & $\boldsymbol{36.27 \times 10^7}$  \\
        \hline
        $\rm Pr$   & 1.0                         & 1.0              \\  
        \hline
        Grid & $864 \times 432$                     & $1200 \times 600 $                     \\
        \hline
        $\nu$ & $6.67 \times 10^{-4}$ & $6.67 \times 10^{-4}$ \\
        \hline
        $\rm{Re}$  & 2091 & 6092 \\
        \hline
        $\langle \rm{Nu} \rangle $  & 24.835 & 38.802 \\
        \hline
        $\eta_\kappa$ & 2.12 & 1.75 \\
        \hline
        $T_d$ & 0.025 & 0.015 \\
        \hline
        $f$ & 164 & 1130 \\
        \hline
    \end{tabular}
    \caption{The parameters used for the reference flows. Frequency of sampling $f$ is in units of probes measurements per turnover time $\tau_0$ while all other parameters are in simulation units. $\rm{Re}$ is defined as ${\rm{Re}} = u_{\rm{rms}} L_z / \nu$ where $u_{\rm{rms}}$ is the root mean square velocity of the flow.}
    \label{tab:Ref_params}
\end{table}

We define two important spectral quantities to understand the small-scale structure of the flow better. First, the time-averaged spectrum (all spectra in this study are measured only in the horizontal direction in a narrow band about the line $z = z_0 = L_z/2$) of the velocity field $\uu$, or the kinetic energy spectrum, given by 
\begin{equation}\label{eq:u-spec}
    E_{\uu}(k_x) = \frac{1}{2} \left\langle \left |\hat{\uu} (k_x,z_0,t) \right|^2 \right\rangle_t,
\end{equation}
where $\hat{\uu} (k_x,z_0,t)$ are the Fourier coefficients of the field $\uu$ and $\langle . \rangle_t$ denotes the time averaging. Similarly, the spectrum of the temperature field is defined as 

\begin{equation}\label{eq:T-spec}
    E_{T}(k_x) = \left \langle  |\hat{T} (k_x,z_0,t) |^2 \right \rangle_t,
\end{equation}
where $\hat{T}(k_x,z_0,t)$ are the Fourier coefficients of the field $T$. 

The magnitude of the problem can be appreciated by looking at the reference flow characteristics, as shown in Fig.~\ref{fig:Refs_comp} where we present $E_\uu$, $E_T$ and instantaneous horizontal cuts of the vertical velocity at $z=z_0$, that is, $v(x,z_0,t)$ for both reference flows. We see in panel (b) that the spectrum of the temperature field becomes steeper when increasing Rayleigh number, indicating the presence of less well defined temperature plumes when turbulence increases. The slope of the temperature spectrum for the moderate $\Ra$ case shows that the temperature field is better correlated over a broader range of scales compared to the higher $\Ra$ flow. The energy spectra for both flows are close to a power law as shown in panel (a). On closer inspection, one can see that the spectrum for the lower $\Ra$ flow is indeed flatter much like the temperature spectrum. A large contrast can also be seen in the energy contained in the first few Fourier modes, where for the lower $\Ra$ flow, the first mode contains far greater energy compared to the successive modes. In the higher $\Ra$ flow the first mode is still the most energetic, but the successive modes still contain a significant amount of energy. This contrast is borne out more clearly in the horizontal cuts of the vertical velocity for both the flow. In the lower $\Ra$ flow (panel (c)), even the instantaneous velocity field is smooth and highly structured with a regular, large-scale pattern with only small fluctuations whereas for the higher $\Ra$ flow (panel (d)), the velocity field in the bulk is far more chaotic and rugged, with a large-scale flow pattern not immediately discernible. 
\subsection{Nudging experiments}

Once $T_n$ is obtained, another flow is initialised with velocity $\UU(\br,t) =0$ and  temperature $\mathcal{T}(\br,t) = 0$ everywhere and evolved according to equations \eqref{eq:R-B-nudged} -- \eqref{eq:R-B-nudged1}. As already described, the thermal boundary conditions at the top and bottom walls are set to a no-thermal flux boundary condition so that the only energy inputs into the nudged system arise from the nudging term. The case of $k_l = 1$ is a special case scenario where it is assumed that the temperature data is available on every point at every time-step. In this case we use the fixed temperature boundary conditions of the Rayleigh-B\'enard equations and we set
$T_{n}(\br,t) = T(\br,t)$ everywhere. 

The nudged simulations are evolved until they attain a statistically stationary kinetic energy. All measurements and further analysis are made in this stationary state. Corresponding to each reference flow, several nudging experiments are performed by varying $k_l$, $\alpha_0$ and  $\chi$ (see Table I and Appendix C). Identically to the spectra already defined for the reference flows (eqns. \eqref{eq:u-spec} - \eqref{eq:T-spec}), we define the kinetic energy spectra $E_{\UU}$ and the thermal energy spectra $E_{\TT}$ for the nudged simulations as well. 

The results presented in the rest of the study are for parameters listed in Table. \ref{tab:Recons_params}. Varying these parameters does not change the qualitative results greatly and an overview of the behaviour of the system on changing $\alpha_0$ and $\chi$ can be found in Appendix C. In the main text we focus on the effects of varying $k_l$ and how these are affected by varying the Rayleigh number of the reference flow. The domain-averaged kinetic energy $K(t)$ of the flow is given by 

\begin{equation}
    K(t) = \frac{1}{2} \langle |\uu(\br,t)|^2 \rangle_V,
\end{equation}

where $\langle \cdot \rangle_V$ denotes the average over the entire domain at a given instant of time. The evolution of the kinetic energy for the lower $\Ra$ reference flow and the higher $\Ra$ reference flows along with those of the reconstructions for different $k_l$ with the parameters from Table. \ref{tab:Recons_params} is shown in Figure \ref{fig:ke_evol}. It is clear that the kinetic energy of the reference flows (black curves) has attained a stationary kinetic energy. The vertical red lines correspond to the time when the reconstructed flows attain a statistically stationary kinetic energy. 

\begin{table}
    \centering
    \begin{tabular}{|>{\centering}p{0.075\textwidth}|>{\centering}m{0.04\textwidth}|>{\centering}m{0.04\textwidth}|>{\centering}m{0.04\textwidth}|>{\centering}m{0.075\textwidth}|>{\centering}m{0.04\textwidth}|>{\centering}m{0.04\textwidth}|>{\centering\arraybackslash}m{0.04\textwidth}|}
        \hline
        \multicolumn{8}{|c|}{\textbf{Reconstructions}}                        \\
        \hline
        \multicolumn{4}{|c|}{$\boldsymbol{\rm{Ra} = 7.206 \times 10^7}$} &  \multicolumn{4}{|c|}{$\boldsymbol{\rm{Ra} = 36.27 \times 10^7}$}\\
        \hline
         $\boldsymbol{N_p}$  & $\boldsymbol{k_l}$  & $\boldsymbol{\chi}$ & $\boldsymbol{\alpha_0}$ & $\boldsymbol{N_p}$ &  $\boldsymbol{k_l}$ & $\boldsymbol{\chi}$ & $\boldsymbol{\alpha_0}$ \\
        \hline
         $3.7 \times 10^5$ & 1 & - & 0.01 & $7.2 \times 10^5$ &   1 & - & 0.01 \\
        \hline
         7688 & 1/7 & 6 & 0.01 & 14792 &   1/7 & 6 & 0.08 \\
        \hline
         3872 & 1/10 & 6 & 0.01 & 7442 &   1/10 & 6 & 0.08 \\
        \hline
         1922 & 1/14& 6 & 0.01 & 3698 &   1/14  & 6 & 0.08\\
         \hline
         800 & 1/22 & 6 & 0.01 &  1485 & 1/22  & 6 & 0.08\\
         \hline
         392 & 1/31 & 6 & 0.01& 735 & 1/31   & 6 & 0.08\\
         \hline 
         162  & 1/48 & 6 & 0.01& 300 &  1/49  & 6 & 0.08\\
         \hline
          81 & 1/68 & 6 & 0.01 & 150 & 1/69 & 6 & 0.08\\
         \hline
          40 & 1/97 & 6 & 0.01 & 72 & 1/100 & 6 & 0.08\\
         \hline
          - & - & - & - & 36 & 1/141 & 6 & 0.08\\
         \hline
    \end{tabular}
    \caption{The parameters used for the reconstructed (nudged) flows in simulation units.}
    \label{tab:Recons_params}
\end{table}

\subsection{Error quantification}
\begin{figure*}
    \begin{overpic}[width = \linewidth]{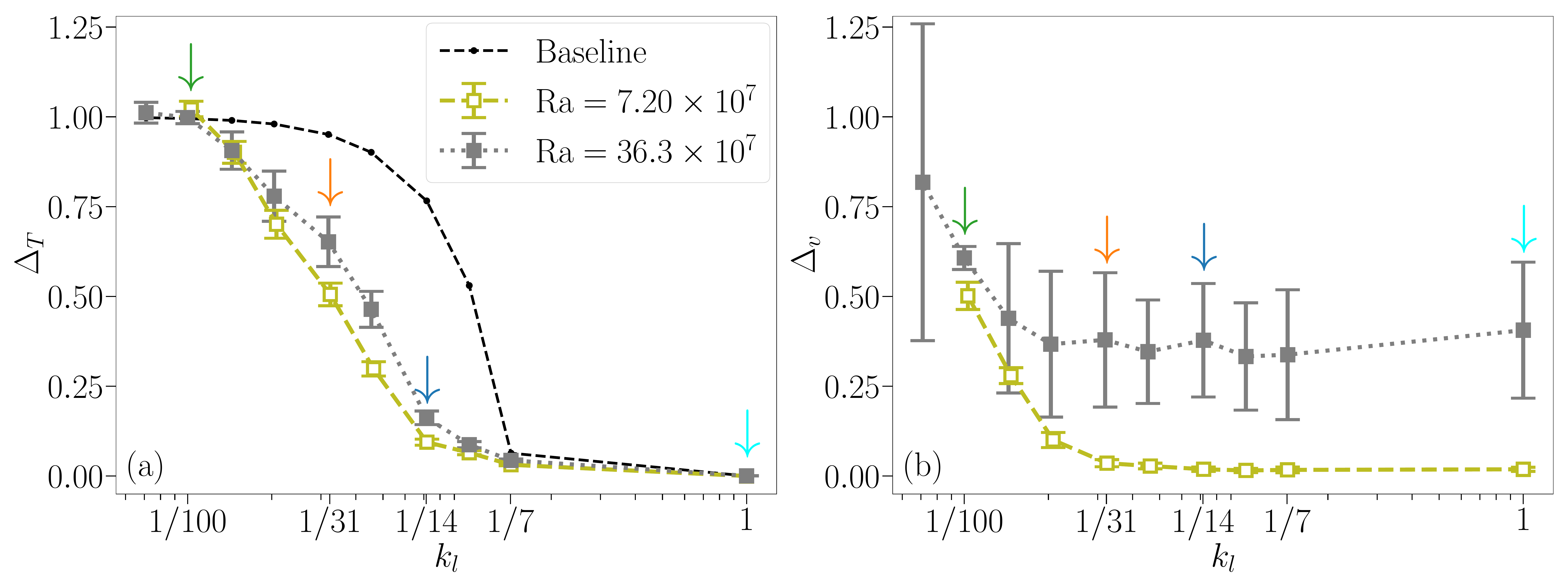}
    \end{overpic}
    \caption{The global reconstruction errors (a) $\Delta_T$, and (b) $\Delta_v$ for reconstructions of the two reference flows are varying $k_l$. Also shown is the baseline reconstruction error (black) for the higher $\Ra$ flow. The arrows indicate the value of $k_l$ presented in detail in this study and whose corresponding nudging fields $T_n$ for the moderate $\Ra$ case are shown in Figure \ref{fig:Nudging_fields}. }
    \label{fig:l2err_global}
\end{figure*}

To compare the reconstructed configurations from the nudging experiments with the ground-truth, we use various measures to quantify the efficacy of the nudging technique applied here. First, we define a point-wise error for temperature and vertical velocity given by 

\begin{equation}
    T_\Delta(\br,t) =  \mathcal{T}(\br,t) - T(\br,t),
\end{equation}
and
\begin{equation}
    v_\Delta(\br,t) =   \mathcal{V}(\br,t) - v(\br,t).
\end{equation}

Next, we define the global L2-error on the temperature and vertical velocity respectively given by 

\begin{equation}
    \Delta_T  = \frac{\langle T_\Delta^2(\br,t) \rangle}{\langle T^2(\br,t) \rangle},
\end{equation}
and
\begin{equation}
    \Delta_v = \frac{\langle v_\Delta^2(\br,t) \rangle}{\langle v^2(\br,t) \rangle},
\end{equation}

where $\langle \cdot \rangle$ indicates the average over the entire domain and the entire (stationary) run-time. We note that perfect synchronisation would lead to a value of $\Delta \sim 0$ while if the reconstructed flow is statistically correct but uncorrelated with the ground truth, we would have $\Delta \sim 2$. $\Delta_T$ for the reconstructed flows is compared with a ``Baseline Reconstruction" where we calculate $\Delta_T$ by assuming $\TT(\br,t) = T_n(\br,t)$ for $\br \in S $, and setting uniformly $\Delta_T = 1$ otherwise - that is setting the temperature uniformly identical to the nudged field where the flow is nudged and equal to the mean temperature, $\TT(\br,t)=0$, every where else. 
\begin{figure*}
    \centering
    \begin{overpic}[width = \linewidth,keepaspectratio]{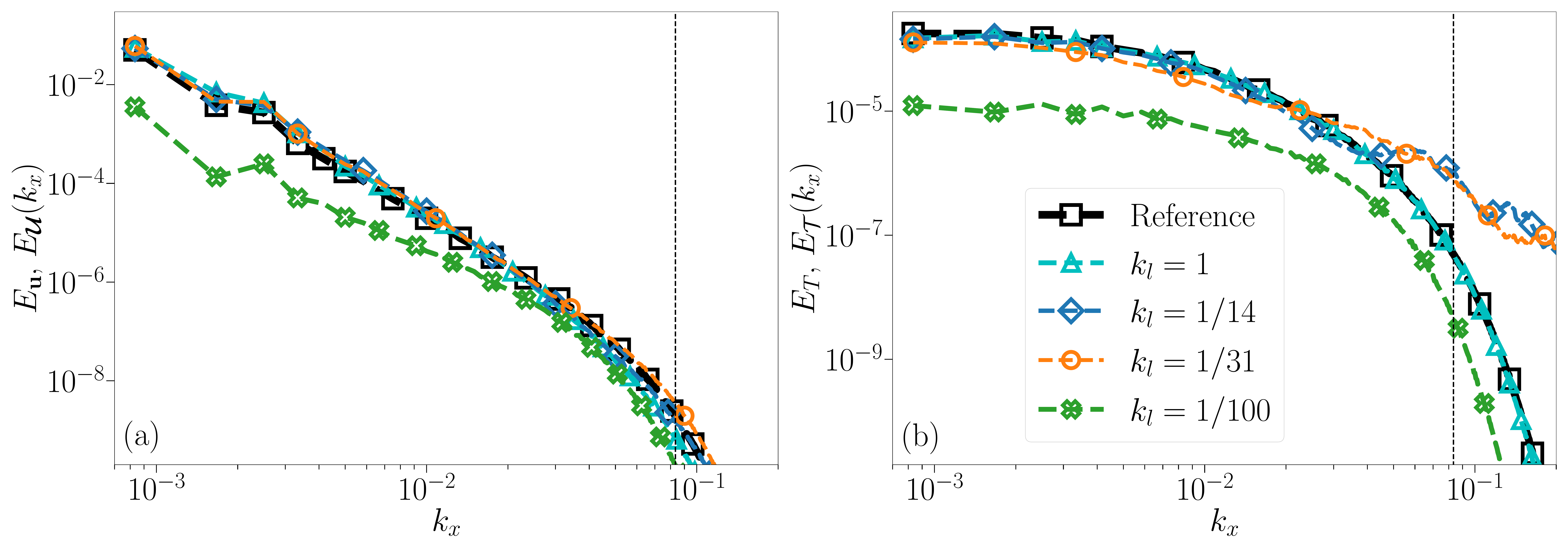}
    \end{overpic}
    \caption{Energy and temperature spectra for the higher $\Ra$ reference flow with $\Ra = 36.3 \times 10^7$ and the reconstructions at various $k_l$ - (a) $E_{\uu}(k_x)$ (black) and $E_{\UU}(k_x)$. (b) $E_T(k_x)$ (black) and $E_{\TT}(k_x)$. The dotted vertical line shows the wavenumber corresponding to the length $\chi$ of the nudging squares.}
    \label{fig:spectra_nudged}
\end{figure*}

To visualise the local errors at each instant, we also define an instantaneous L2-error on the temperature and vertical velocity given by 

\begin{equation}\label{eq:l2inst_T}
\delta_T(\br,t) = \frac{ T^2_\Delta(\br,t)}{\langle T^2(\br,t) \rangle_{x,t}},
\end{equation}

\begin{equation}\label{eq:l2inst_v}
\delta_v(\br,t) = \frac{ v^2_\Delta(\br,t)}{\langle v^2(\br,t) \rangle_{x,t}}
\end{equation}

respectively where $\langle \cdot \rangle_{x,t}$ indicates the time and spatial averages at a given height $z$.

For a comparison of the scale-by-scale reconstruction, we also define the spectrum of the errors of temperature and vertical velocity respectively given by 

\begin{equation}
    E^\Delta_T(k_x) =  \left \langle |\hat{T}_\Delta (k_x,z_0,t) |^2 \right \rangle_t,
\end{equation}
and
\begin{equation}
    E^\Delta_v(k_x) =  \left \langle |\hat{v}_\Delta (k_x,z_0,t) |^2\right \rangle_t,
\end{equation}

where $\hat{T}_\Delta (k,t) $ and $\hat{v}_\Delta (k,t)$ are respectively the fourier coefficients of the fields $T_{\Delta}$ and $v_\Delta$. 


\section{Results}\label{sec:Results}

\subsection{Instantaneous and Global L2-Errors}

We start by showing visualizations of the reference temperature field $T$, the nudged temperature field $\TT$ and the temperature error field $\delta_T$ at one given instant for both the flows in Fig.~\ref{fig:Confs_comp_T} and likewise for $v$, $\mathcal{V}$ and $\delta_v$ in Fig.~\ref{fig:Confs_comp_v}. Further, the vertical time-averaged profile of the errors, that is $\langle \delta_T \rangle_{x,t}$ and $\langle \delta_v \rangle_{x,t}$ are also overlaid on the snapshots of the instantaneous errors (right-most panels, solid yellow line). The nudged fields correspond to a case with $k_l=1/14$, i.e., a high density of probes as seen in Fig.~\ref{fig:Nudging_fields}(b). Both reference flows are characterised by a rising hot plume and a falling cold plume and two large-scale counter-rotating vortices. The visualisations show clearly that the higher $\Ra$ flow (bottom row) shows an abundance of fine-scale structure and further, the plumes are poorly-defined, in contrast to the moderate $\Ra$ flow, where while the flow is not completely laminar, the plumes are clearly demarcated. The nudging protocol is able to accurately reconstruct the temperature field, even at higher Rayleigh number. In this case the maximum error is of the order of $10\%$ and is concentrated on the plume and vortex structures while the thermal boundary layers show larger errors as clear from the average profile. This profile for the moderate $\Ra$ case is smooth and the value of the error doesn't change very sharply depending on whether the particular region is nudged or not (that is, whether $\br \in S$ or not). In contrast, in the reconstruction of the higher $\Ra$ flow, the profile of the error shows highly oscillatory behaviour, indicating that while errors are lower in the nudged regions, they are relatively higher even in adjoining regions which are not nudged. This is a manifestation of the ruggedness of the temperature field when the flow is more turbulent. This is an indication that the low magnitude of error is solely due to the high density of probes in the system and not due to a complete synchronization of the reconstruction with the reference flows. Closer inspection of this averaged profile (yellow curves) near the top and bottom walls for both the flows shows that the error reaches the highest value close to the thermal boundaries. The large error near the thermal boundary is due to the relatively small thickness of the thermal boundary layer and the steep temperature gradient away from the wall. It is clear that to capture the precise behaviour of the temperature near the boundary walls needs a higher density of probes immersed near the walls.

\begin{figure*}
    \centering
    \begin{overpic}[width = \linewidth]{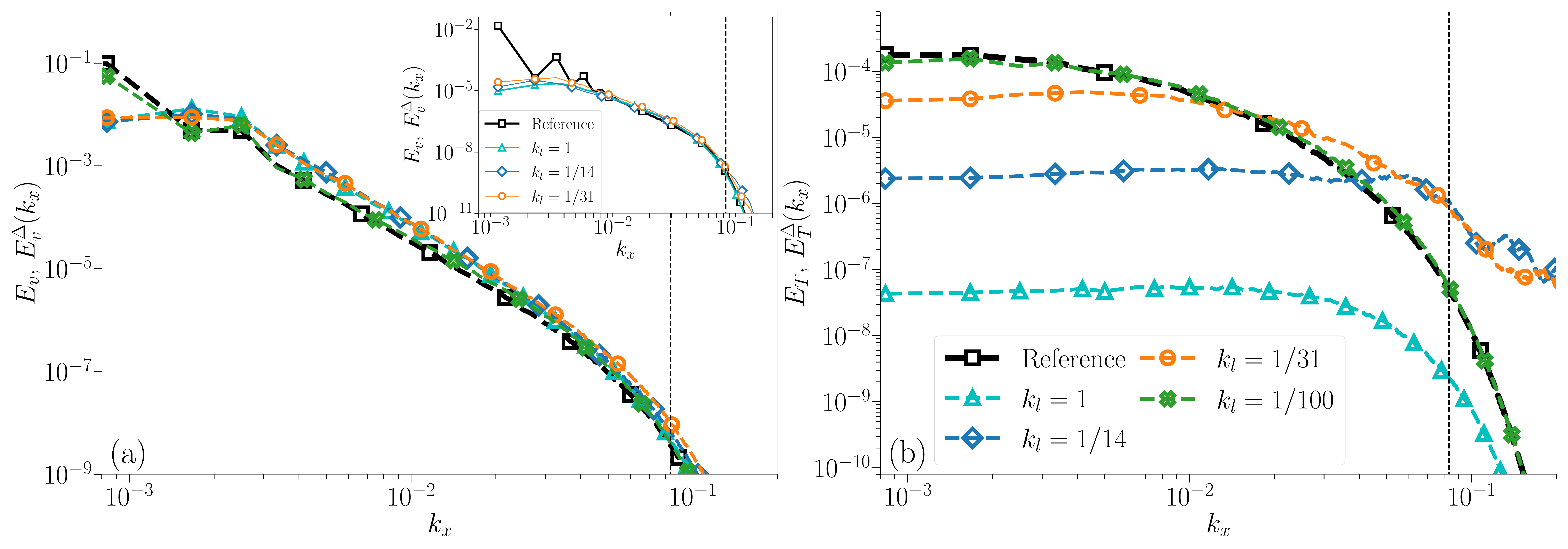}
    \end{overpic}
    \caption{The spectra of the errors (a) $E_v^\Delta$ and (b) $E_T^\Delta$ for various $k_l$ for reconstructions of the higher $\Ra$ flow with $\Ra = 36.3 \times 10^7$. For comparison, the reference spectra (a) $E_v$ and (b) $E_T$ are also shown (black squares). The inset in panel (a) shows $E_v$ for the moderate $\Ra$ reference flow with $\Ra = 7.2 \times 10^7$ and $E^{\Delta}_v$ for the corresponding nudging experiments. The dotted vertical line shows the wavenumber corresponding to the length $\chi$ of the nudging squares. }
    \label{fig:error_spectra}
\end{figure*}

The situation is different for the velocity field, which is reconstructed indirectly through the thermal forcing and depends on the response of the velocity field to the temperature field. We see from the reference velocity fields (Panels (a) and (d) in Fig. \ref{fig:Confs_comp_v}), the regions with rising fluid (red) and falling fluid (blue) are more clearly delineated in the lower $\Ra$ flow unlike in the higher $\Ra$ flow. The nudged field is in good agreement with the truth for the lower Rayleigh number case (top row) while large systematic errors on the whole volume develop for the higher $\Ra$ flow (bottom row) where the reconstructed velocity fails to capture the precise instantaneous shape of the reference while still broadly capturing the regions of rising and falling fluid accurately. The average profile of the errors for both cases are smooth. The invisibility of the nudging squares to the velocity field is a clear indication that the velocity field is not set locally by the local temperature but rather by the large-scale interplay between the temperature and velocity field. The reconstruction of the vertical velocity is most accurate in the bulk. Notice that due to the no-slip boundary condition ($\uu = 0$) imposed at the top and bottom walls for the reference as well as reconstructed flows, the particular form of L2-error chosen here ceases to be well-defined at the vertical boundaries. 

In Fig.~\ref{fig:l2err_global} we show the main quantitative summary of our study, where we plot $\Delta_T$ and $\Delta_v$ as a function of $k_l$ for the reconstructions of both flows. As seen in panel (a), the reconstruction error for temperature decreases when increasing the number of probes. A transition to synchronization-to-data can be seen at $k_l \sim 1/14$, corresponding to a typical wavenumber $k_x \sim 0.036$ or $1/28$ in Fig.~\ref{fig:Refs_comp}. Equivalently, this corresponds to providing information at a separation of $\sim 7 \eta_\kappa$ for the moderate $\Ra$ flow and $\sim 8 \eta_\kappa$ for the higher $\Ra$ flow. The transition to {\it perfect} synchronization of the temperature fields occurs at similar scales in the two flows studied, and is similar to that observed in homogeneous and isotropic flows \cite{Lalescu13,clark_di_leoni_synchronization_2020}. The baseline reconstruction error approaches $0$ for $k_l = 1/7$, the scale at which for the chosen $\chi$, the nudging squares cover the entire domain and thus, the baseline error includes only the small-scale temperature errors introduced by the application of a constant nudging temperature within each nudging square. Setting $\chi$ to be larger would cause these small-scale errors to be far larger while setting $\chi$ too small would lead to a reconstructed flow with significantly smaller kinetic energy and increase the effective distance between the probes, thus leading to larger global errors. The behaviour for changing $\chi$ for a given $k_l$ is discussed in Appendix C. 

As for the reconstruction of the velocity field shown in panel (b), while the velocity field can be accurately reconstructed in the lower Rayleigh number case, the higher Rayleigh number case does not synchronize even for full information on the temperature ($k_l=1$), showing a plateau at close to $40\%$ for the average minimum error committed, the plateau being reached already at $k_l=1/48$. The result is not completely surprising, indicating that for high Rayleigh number the {\it slaving} of the velocity field is less and less effective, and the imposition of more and more information on the forcing field (here the temperature) is not sufficient to recover full synchronization when turbulence is intense enough. 

\begin{figure*}
    \centering
    \begin{overpic}[width = \linewidth]{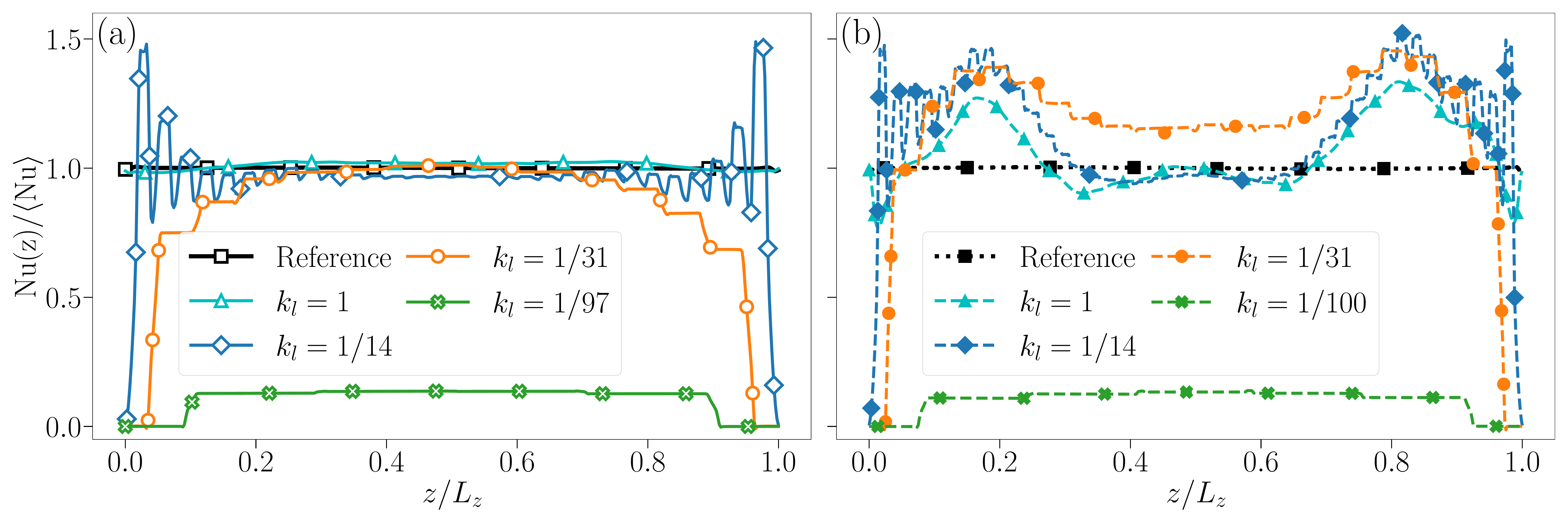}
    \end{overpic}
    \caption{Profile of the Nusselt number for the ground truth and the reconstructions  of the reference flow for (a) moderate $\Ra$ flow with $\Ra = 7.2 \times 10^7$ and (b) higher $\Ra$ flow with $\Ra = 36.3 \times 10^7$. The reference Nusselt number $\rm{Nu}(z)$ and the reconstructed Nusselt number $\textit{Nu}(z)$ are divided by the corresponding reference Nusselt number $\langle \rm{Nu} \rangle$ with magnitude (a) $\langle \rm{Nu} \rangle = 24.835$ and (b) $\langle \rm{Nu} \rangle = 38.802$ respectively. }
    \label{fig:Nusselt}
\end{figure*}

\subsection{Scale-by-scale spectral properties and spectral errors}

In Figure \ref{fig:spectra_nudged} we show the statistical reconstruction of the flow properties by plotting the spectra for the reconstructed higher $\Ra$ flows - $E_{\UU}$ and $E_\TT$  - the same quantities as shown in the first two panels of Fig. (\ref{fig:Refs_comp}) - for various $k_l$ along with the ground truth in black. Panel (a) shows that the nudging experiments retrieve a flow with a velocity field with similar dynamic properties as the reference at all scales. In panel (b) showing the spectrum of the thermal energy, as expected the case with $k_l = 1$ follows the ground truth exactly while the case with $k_l = 1/14$ captures the correct large-scale dynamics and at the same time introduces spurious correlations at smaller scales. This is due to the finite sizes of the nudging squares within which the nudging temperature is imposed uniformly, leading to systematic errors of small magnitudes for points close to $\bX^i$. 

Further insight into the scale-wise behaviour is provided in Fig.~\ref{fig:error_spectra}, which shows the error spectra $E_v^\Delta$ and $E_T^\Delta$ and the corresponding reference spectra $E_T$ and $E_v$. The scale-wise relative error can be gauged by the vertical displacement between the reference spectrum (black) and the reconstructed spectra. In panel (b) the error spectrum is nearly flat at large-scales with a value strongly dependent on $k_l$ while $E_T$ falls off gradually, indicating an increasing relative error for increasing $k_x$ with the lowest relative errors at the largest scales. For larger $k_x$ (smaller length-scales) we see a clear manifestation of the errors introduced by the finite nudging squares for $k_l = 1/14$ and $k_l = 1/31$. The error-spectrum for $k_l = 1/100$ is identical to the temperature spectrum of the ground truth, a result of the fact that the probes are extremely sparse and hence the reconstructed temperature field is nearly $0$ everywhere. The black curve in panel (a) is the spectrum of the vertical velocity $E_v(k_x)$. The relative error is smallest at the largest scale and at smaller scales (larger $k_x$), we see that the error spectra and reference spectrum decay almost identically and hence shows a nearly constant relative error. It is in this aspect that we see the greatest contrast between the two $\Ra$ flows. The inset shows that reconstructions of the moderate $\Ra$ flow has far smaller relative error at the largest scales and this even persists for larger values of $k_x$. The small-scale behaviour is similar to what is observed for the higher $\Ra$ case in the main panel. 

Another important point to note is that the value of the error spectrum $E_v^\Delta$ shows little dependence on $k_l$ and that the only scales of the velocity field that truly synchronize are the largest ones. The largest scale structure present are the hot and cold thermal plumes, which are captured correctly by the reconstruction. When the degree of turbulence is lower, there is a relative absence of fine-scale structures in the flow and most of the energy is contained in the large-scale alone - this large-scale synchronisation persists at some mid-range and smaller scales as well. For more turbulent flows where the plumes are less well-defined and the velocity field is more rugged, the synchronisation at the largest scales exists but to a smaller degree and it does not persist at smaller scales. This leads eventually to a much larger global error. These observations are explored in further detail below.

\subsection{Inferring Nusselt Number}

The Nusselt number defined in equation (\ref{eq:Nu-defn}) measures the heat transfer due to convection relative to that due to conduction in the Rayleigh-B\'enard system. Similarly, for the reconstructed flows we define the Nusselt number as 
\begin{equation}
    \textit{Nu}(z) = \frac{\mathcal{V} \TT - \kappa \partial_z \TT}{\frac{\kappa \Delta T }{L_z}},
\end{equation}
where $\Delta T$ is the temperature difference between the horizontal walls for the corresponding reference flow. As already noticed, given only a set of temperature measurements it would be impossible to infer neither the Rayleigh number nor the Nusselt number.  In Fig. \ref{fig:Nusselt} we show the average profile of the Nusselt number as a function of the distance from the wall for the reference as well as the reconstructed flows for the moderate $\Ra$ flow and the higher $\Ra$ flow in panels (a) and (b) respectively for different $k_l$. As expected, for the moderate Rayleigh number flow, the Nusselt number is perfectly reconstructed as soon as we are close to the transition to perfect synchronization, except at the walls where we force $\rm{Nu} \to 0$ by setting an adiabatic boundary condition. In the $k_l = 1$ case where the walls have fixed temperature identical to the reference, the Nusselt profile is reconstructed perfectly. In the higher $\Ra$ case on the other hand, even when the temperature field is perfectly synchronised, there is a large discrepancy between the reconstructed and the reference values. It is interesting to notice that the main source of errors comes from the region just after the thermal boundary layer, where the correlation between plumes and vertical velocity drafts are particularly important. In the bulk the reconstruction is accurate when the probes reach a high enough density ($k_l \geq 1/14$). 

\subsection{Varying Rayleigh Number}

\begin{figure*}
    \centering
    \begin{overpic}[width = \linewidth]{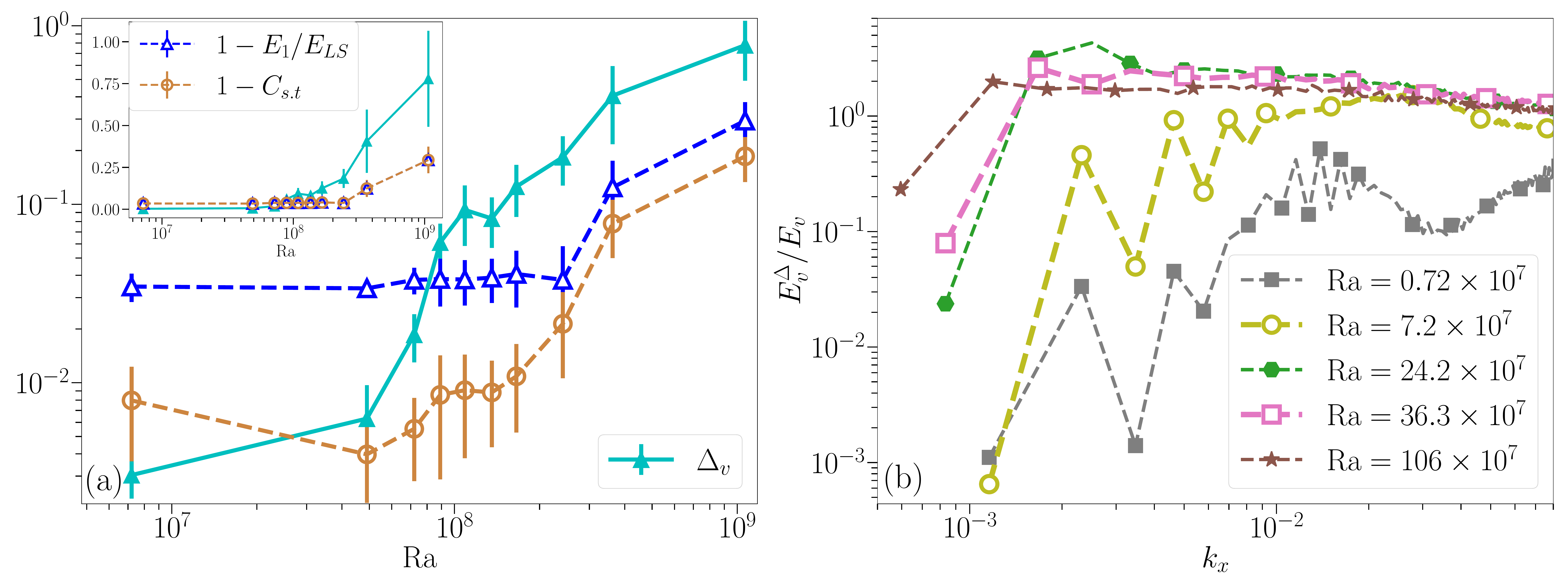}
    \end{overpic}
    \caption{(a) $\Delta_v$ for the reconstructed flows with $k_l = 1$ as a function of $\Ra$ of the reference flows (cyan, filled triangles) along with $1-E_1/E_{\rm{tot}}$ and $1 - C_{s.t}$for the reference flows. Inset shows the same plot without the log-scale on the y-axis (b) Relative scale-wise error $E_v^\Delta/E_v$ for the reconstructed flows. }
    \label{fig:Ra_var}
\end{figure*}

To further corroborate our findings reported thus far regarding the large-scale synchronization of the reconstructed flows with the reference flows, we conduct another series of nudging experiments for varying values of $\Ra$ with $k_l = 1$, that is, complete information on temperature. For all the flows, we have $\Delta_T \lesssim 10^{-3}$. We show the corresponding values of $\Delta_v$ as a function $\Ra$ in Fig.~\ref{fig:Ra_var}(a). The inset of the Figure shows the same but in the semilog scale. In Fig.~\ref{fig:Ra_var}(b) we show the spectra of the errors $E^\Delta_v$ compensated by the spectra of the reference $E_v$ for a selection of cases. At $\Ra \sim 10^7$ the velocity field is reconstructed correctly by the nudging protocol, but at around $\Ra \sim 10^8$ a transition occurs, and the nudged flow is not able to synchronize to the reference flow anymore. 

In order to get a better understanding of this phenomenon, we calculate the strength of the Large Scale Circulation (LSC) by looking at the ratio of the energy in the lowest wavenumber (first Fourier mode) $E_1$ and the energy $E_{\rm{LS}}$ contained in the first four Fourier modes \cite{RBflowmodes} given by

\begin{equation}
    E_{\rm{LS}} = E_1 + E_2 + E_3 + E_4,
\end{equation}

where $E_i$ is the average energy contained in the $i$-th Fourier mode. In Fig.~\ref{fig:Ra_var}(a), we plot (blue triangles) the deviations from the LSC strength for the various reference flows, i.e. $1-E_1/E_{\rm{LS}}$. At low $\Ra$ most of the energy is contained in the largest mode, something that is suggested by Fig.~\ref{fig:Refs_comp}(c), while when $\Ra$ is increased, the flow becomes more turbulent and disordered, as seen in Figs.~\ref{fig:Refs_comp}(a) and (d) and the energy contained in the smaller scales is more and more significant. We see that the deviations from the LSC scale similarly as $\Delta_v$. The results are qualitatively the same even if we consider $E_{\rm{LS}}$ as the sum of all Fourier modes of the energy instead of just the first four, with the mean of $1 - E_1/E_{\rm{LS}}$ merely shifted upward. 

Further, we consider a continuous saw-tooth function $\lambda$ with period $L_x$ and translated by a distance $a$ given by

\begin{equation}
    \lambda(x,a) = \begin{cases}
    \frac{4}{L_x}(x-a), & 0 \leq x \leq \frac{L_x}{4}+a \\
    - \frac{4}{L_x}(x-a) + 2, & \frac{L_x}{4}+a < x \leq \frac{3L_x}{4} -a \\
    \frac{4}{L_x} (x-a) -4, & \frac{3L_x}{4} -a < x \leq L_x
    \end{cases}
\end{equation}
with $a$ chosen at each instant such that $$\int_{x=0}^{x=L_x} v(x,L_z/2,t) \cdot \lambda(x,a) dx$$ is maximum. The similarity between the saw-tooth function and the vertical velocity signal is then calculated as the time-averaged Pearson Correlation Coefficient, denoted $C_{s.t}$. In  Fig.~\ref{fig:Ra_var}(a) we also show the value $1-C_{s.t}$ which quantifies the deviations from the saw-tooth mode for the vertical velocity. We see again that $\Delta_v$ scales similar to the deviations from the saw-tooth function.

This picture indicates that at low $\Ra$ supplying the location of the hot and cold plumes is enough to reconstruct the flow accurately, as the plumes set the structure of the largest scales with a single dominant mode that can be approximated as a cosine or a saw-tooth. When the flow becomes more turbulent the number of structures that can exist in the flow increases and, thus, information on the plumes positions alone cannot uniquely determine the proper solution.
This, further coupled with the fact that there exist several features of the flow at smaller scales leads to a poor reconstruction of the Rayleigh-B\'enard Convection at larger Rayleigh numbers. 

\section{Conclusion}\label{sec:Concl}

We have presented here a study of nudging applied to a thermally driven fluid in the presence of a bulk external forcing (gravity), where the only control variable is the temperature. We have varied the quantity of information as well as the quality of information to be used. We have shown that given a greater quantity of information on the temperature, the nudging method used here yields a superior reconstruction of the temperature field, with a transition to full-synchronisation around $k_l \sim 0.07$ corresponding to a distance between probes of around $7 \eta_\kappa$ for both flows. The reconstruction of the velocity field on the other hand saturates from a relatively lower $k_l$ and supplying more information, even the full temperature field with near-perfect reconstruction of temperature fails to further improve the velocity reconstruction. The degree of synchronisation between the reconstructed and the reference velocity fields depend on the Rayleigh number and the degree of turbulence in the reference flow, with the largest scales of the flow synchronising the most effectively while at smaller scales, the velocity fields remain largely asynchronous. 

The quality of information was varied by following a Lagrangian approach as well as an Eulerian approach and in the construction of the nudging field $T_n$ given the information on the temperature. While the different approaches yielded slightly different reconstructions of the temperature field, the accuracy of the reconstruction of the velocity field remains nearly the same. 

Rayleigh B\'enard convection is driven by the response of the vertical velocity to the local temperature. The correlation between these two fields however manifests first at the largest scales in the velocity and local temperature measurements alone fails to predict the local velocity. Providing sufficient information on the temperature provides information of these largest scale features - the plumes - and hence the large-scale flow. Our study thus helps to understand the fundamental role played by the temperature field alone in the Rayleigh-B\'enard convection and answers the question – how much does the velocity field depend on the quantity and quality of information available on the temperature field? The accurate reconstruction of the velocity field at lower Rayleigh numbers is a result of the relative absence of smaller and fine scale structure in such flows. At higher $\Ra$, when the thermal flow is more turbulent, richer in small-scale structures, with thermal plumes less well-defined and rapidly fluctuating velocity field, the indirect reconstruction of the velocity field reproduces the correct flow dynamics and synchronises relatively poorly even at the largest scale, though the relative error on the first Fourier mode is still of the order $\sim 10^{-1}$ for the flow with ${\Ra} \sim 10^9$ and highly turbulent. To accurately reconstruct the smaller and smaller scales would require additional inputs into the system - it is conceivable, for example, that even a small amount of data on the velocity field could drastically improve the accuracy of the reconstructed velocity field and that there exists an optimum way to supply a combination of temperature and velocity data. In this study we focus exclusively on the reconstructions using the temperature field alone and leave the possibility of using a combination of temperature and velocity fields for future work. 

\section*{Acknowledgements}
This project has received funding from the European Union's Horizon 2020 research and innovation programme under the Marie Sklodowska-Curie grant agreement No 765048. This work was supported by the European Research Council (ERC) under the European Union’s Horizon 2020 research and innovation programme (Grant Agreement No. 882340).

\section*{Data Availability Statement}
Data available on request from the authors - The data that support the findings of this study are available from the corresponding author upon reasonable request.

\appendix

\section{Lattice Boltzmann Method}\label{sec:AppendixA}
All  fields are obtained by solving the Rayleigh-B\'enard and the corresponding nudging equations using the Lattice-Boltzmann method in 2D with a standard D2Q9 grid. Two sets of populations $f$ and $g$ which represent the fluid phase and the thermal phase respectively are introduced with their time evolution governed by the BGK collision operator. \cite{HEetal_2population} The populations solve the Lattice Boltzmann equations
\begin{equation}
    \label{LB_equations}
    \begin{gathered}
        f_i(\br+\mathbf{c}_i, t + 1) = f_i(\br,t) + \frac{f_i - f^{\rm eq}}{\tau_f}, \\
        h_i(\br+\mathbf{c}_i, t + 1) = h_i(\br,t) + \frac{h_i - h^{\rm eq}}{\tau_h}  
    \end{gathered}
\end{equation}
where the vectors $\mathbf{c}_i $ for $i=1,\dots,9$ are the discrete particle velocities, $\Delta t$ is the lattice time-step, so that $\mathbf{c}_i \Delta t$ go from each lattice point to the 8 nearest neighbouring lattice points in the uniform 2D grid and $\mathbf{c}_0 = 0$. We have the time-step and the grid spacing respectively $\Delta t = \Delta r = 1$, as is the standard practice. $f^{\rm eq}$ and $h^{\rm eq}$ are the equilibrium population distributions as defined in \cite{HEetal_2population} given by 
\begin{equation}
    \begin{gathered}
    f^{\rm eq} = w_i \rho \left ( 1 + \frac{\uu\cdot \mathbf{c}_i}{c^2_s} + \frac{(\uu\cdot \mathbf{c}_i)^2}{2c^4_s} - \frac{\uu\cdot \uu}{2c^2_s}\right ) \label{eq:feq}\\
    h^{\rm eq} = w_i T \left ( 1 + \frac{\uu\cdot \mathbf{c}_i}{c^2_s} + \frac{(\uu\cdot \mathbf{c}_i)^2}{2c^4_s} - \frac{\uu\cdot \uu}{2c^2_s}\right )
    \end{gathered}
\end{equation}
where $w_i$ are the weights for each population set by the grid used, D2Q9 in this study. $c_s = 1/\sqrt{3}$ is the lattice speed of sound set by the choice of $\mathbf{c}_i$. $\tau_f$ and $\tau_h$ are respectively the fluid and the thermal relaxation times which set the values for kinematic viscosity $\nu$ and thermal conductivity $\kappa$ as 

\begin{equation}
    \begin{gathered}
        \nu = c_s^2 (\tau_f - 0.5); \\
        \kappa = c_s^2 (\tau_h - 0.5)
    \end{gathered}
\end{equation}

To account for the buoyancy force term, the Guo-forcing scheme \cite{Paper:Guo-forcing} is employed and the equation for $f_i$ is modified as 

\begin{equation}
    f_i(\br+\mathbf{c}_i , t + 1) = f_i(\br,t) + \frac{f_i - f^{\rm eq}}{\tau_f} + M_i
\end{equation}

with 

\begin{equation}
    M_i = \left( 1 - \frac{1}{2 \tau_f} \right) w_i \left ( \frac{\mathbf{c_i} 
    - \uu}{c^2_s} + \frac{(\mathbf{c_i} \cdot \uu) \mathbf{c_i}}{c^4_s}  \right) \cdot \mathbf{F}
\end{equation}

where $\mathbf{F}$ is the force vector. The fluid hydrodynamic quantities at each point in space and time are obtained from the various moments of the populations as

\begin{align}
    \rho &= \sum_i f_i;  \\
    \uu &= \frac{1}{\rho} \sum_i f_i \mathbf{c}_i + \frac{\mathbf{F}}{2\rho}; \label{eq:LB_uhdydro}
\end{align}
The ease of implementation of the Guo-forcing scheme is from the fact that the velocity $\uu$ that enters the expression for $f^{\rm eq}$ in equation \eqref{eq:feq} is the same as the hydrodynamic velocity obtained in equation \eqref{eq:LB_uhdydro}. This isn't the case for other forcing schemes.

The nudging experiments require the addition of the nudging term (see equation \eqref{eq:T-nudged}) to the temperature evolution equation. The two-population method is ideal for this study since it allows us to set the Prandtl number by varying $\tau_f$ and $\tau_h$ independently and makes the addition of a heat source term (nudging term) in equation~\ref{eq:R-B-nudged} simpler. The equation for $h_i$ is modified as \cite{Thermal_forcing_Seta}

\begin{equation}
    \begin{split}
    h_i(\br+\mathbf{c}_i , t + 1) 
    &= h_i(\mathbf{r},t) + \frac{h_i - h^{\rm eq}}{\tau_h} \\ &+ \left( 1 - \frac{1}{2 \tau_h}\right) w_i  Q 
    \end{split}
\end{equation}

where $Q = - \alpha (\TT - T_{n})$ is the required source term. For the ground truth, $Q = 0$. The temperature is then obtained at each lattice grid point from the thermal populations $h_i$ as 
\begin{equation}
    \TT= \sum_i h_i + \left( 1 - \frac{1}{2 \tau_h}\right)  Q
\end{equation}

\begin{figure}
    \centering
    \includegraphics[width = \columnwidth]{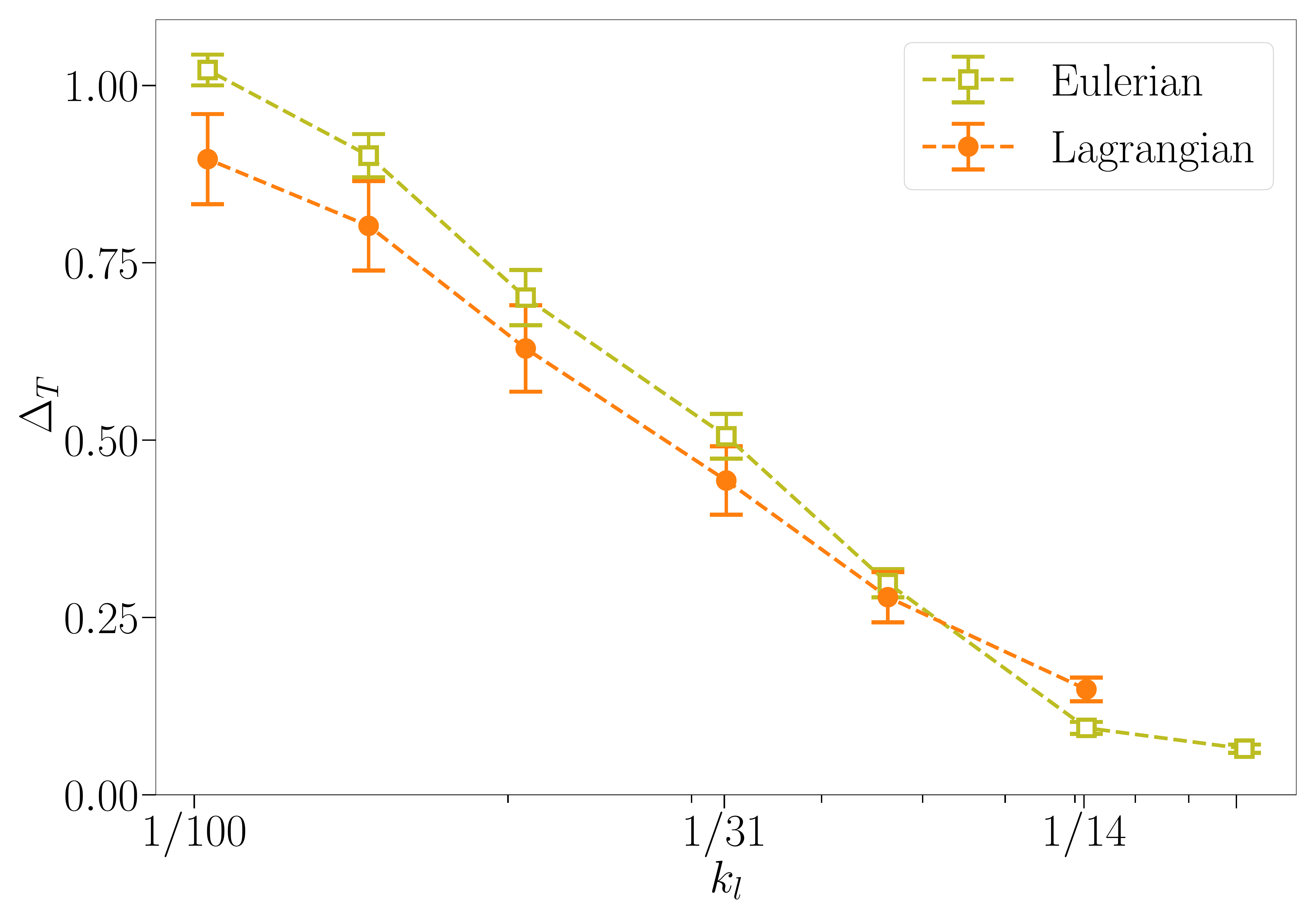}
    \caption{The global reconstruction error $\Delta_T$ as a function of $k_l$ with Eulerian and Lagrangian nudging for reconstructions of the lower $\Ra$ flow with $\Ra = 7.2 \times 10^7$}
    \label{fig:EulvsLag}
\end{figure}

\begin{figure*}
    \centering
    \includegraphics[width = \linewidth]{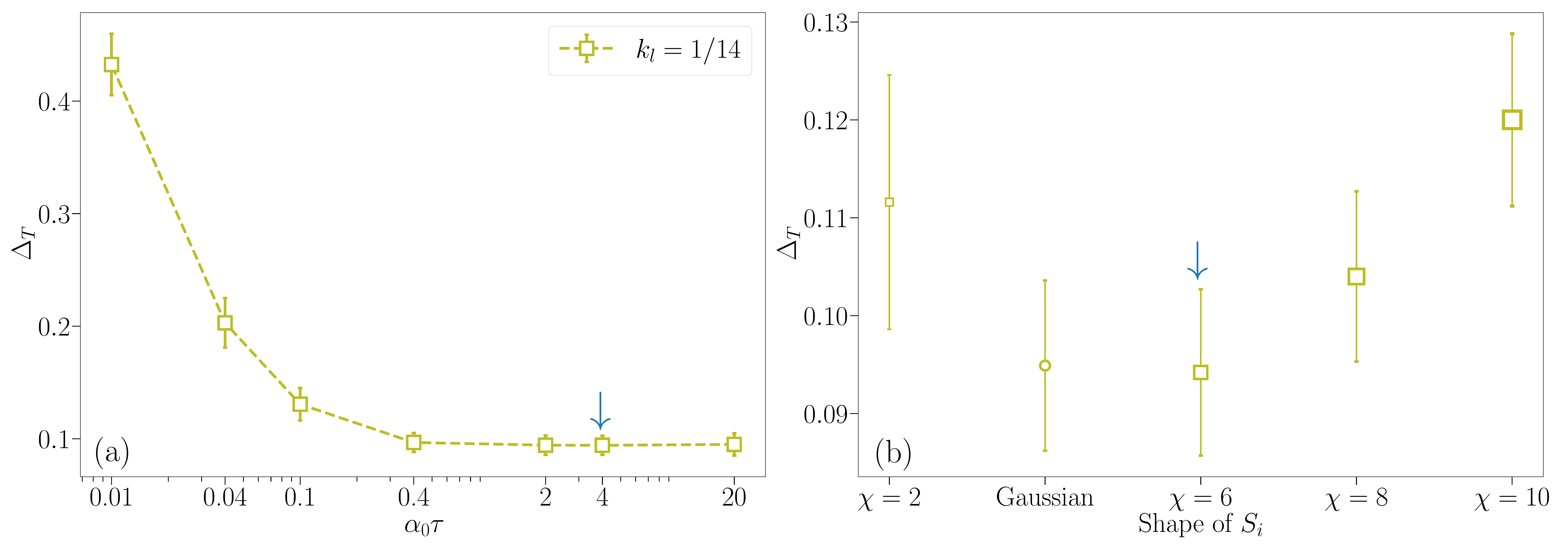}
    \caption{(a) $\Delta_T$ for reconstructions of the lower $\Ra$ flow with $\Ra = 7.2 \times 10^7$ for $k_l = 1/14$ as a function of $\alpha_0 \tau$, where only $\alpha_0$ is varied. (b) $\Delta_T$ for reconstructions of the lower $\Ra$ flow for different configurations of the nudging squares for $k_l = 1/14$ . The blue arrows in both panels indicate the value chosen for more detailed analysis for this study.}
    \label{fig:alpha_squares}
\end{figure*}

\section{Lagrangian Nudging}\label{sec:AppendixB}

Apart from nudging using fixed temperature probes on a uniform grid as shown in figure \ref{fig:Nudging_fields}, another novel approach to nudging is to nudge along the trajectories of tracer particles as in \cite{PatoNudgingN-Seqn}, which in our case is the same as having probes as passive tracer particles. Thus, the probe locations $\bX^i(t)$ for $i=  1,2,\dots , N_p$ satisfy
\begin{equation}
    \frac{d \bX^i(t)}{d t} = \uu (\bX^i(t),t)
\end{equation}
Consecutive measurements of temperature and position are made with the same frequency $f$ as the Eulerian case and both quantities are again interpolated linearly to obtain a reading at each time-step. Off-grid temperatures are obtained using a bi-linear interpolation. The nudging-squares for each probe is constructed with the nearest grid point to the interpolated probe location as the centre. In the regions where the nudging squares of multiple probes intersect, $T_{n}(\br,t)$ is set as the mean of the $\Tp(t)$ of the intersecting probes while $\alpha(\br,t)$ is set to be $n \alpha_0$, where $n$ is the number of probes whose nudging squares intersect at $\br$.

As shown in Figure. \ref{fig:EulvsLag}, the Lagrangian approach to nudging shows only a marginal improvement in reconstruction of the temperature field for smaller $k_l$ when compared to the Eulerian approach while the reconstruction of the velocity field remains unchanged and identical to the Eulerian case for the corresponding $k_l$. 

\section{Selection of the Parameters}\label{sec:AppendixC}

The results presented in the main text have fixed $\chi = 6$ and $\alpha_0 \tau = 4$. These results are representative of results for a larger family of parameters. The main quantitative result of the study - once $\Delta_T$ reaches a low enough value for a given flow, $\Delta_v$ reaches a saturation value - still holds true. Here we report the effect of changing the two parameters, namely the value of $\alpha$ and the size of the nudging squares $\chi$. The results are summarized in Figure \ref{fig:alpha_squares}. In panel (a) we see that $\Delta_T$ is far larger for smaller values of $\alpha_0$ with a transition to a saturation value of $\Delta_T \sim 0.1$ at $\alpha_0 \tau \sim 0.4$. In panel (b) we see the variation of $\Delta_T$ for changing size of the nudging squares $\chi$. Additionally, the ``Gaussian" case refers a nudging experiment where we set 
\begin{equation}
    \alpha(\br,t) = \alpha_0 \exp{\left( -\frac{|\br - \bX^i(t)|^2}{2} \right)}, \qquad \text{for $\br \in S_i$}
\end{equation}

to examine whether the discontinuity of the forcing term has a significant impact on the reconstruction of the temperature field. There exists an optimum value of $\chi$ for the given flow, but the change in $\Delta_T$ is not very significant. On one hand, increasing the size of nudging squares introduces more temperature errors at the smallest scale while on the other hand, very small nudging squares lead to large regions between probes which aren't nudged and hence decorrelate from the known temperature of the probe location. The reconstruction error in velocity for changing $\alpha_0$ as well as changing $\chi$ behaves as anticipated already in figure \ref{fig:l2err_global} - $\Delta_v$ decreases with decrease in $\Delta_T$ upto a certain point and then saturates. This behaviour is qualitatively similar for the higher $\Ra$ flow as well. 

\bibliography{aipsamp}

\end{document}